\documentclass[
 aip,
 amsmath,amssymb,
preprint,%
]{revtex4-1}
\usepackage{xcolor}
\usepackage{graphicx}
\usepackage{dcolumn}
\usepackage{bm}
\usepackage[utf8]{inputenc}
\usepackage[T1]{fontenc}
\usepackage{mathptmx}
\usepackage{etoolbox}

\newcommand{\edit}[1]{\textcolor{black}{#1}}

\makeatletter
\def\@email#1#2{%
 \endgroup
 \patchcmd{\titleblock@produce}
  {\frontmatter@RRAPformat}
  {\frontmatter@RRAPformat{\produce@RRAP{*#1\href{mailto:#2}{#2}}}\frontmatter@RRAPformat}
  {}{}
}%
\makeatother
\begin{document}

\title[Bidirectional Wave-Propelled Capillary Spinners]{Bidirectional Wave-Propelled Capillary Spinners}

\author{Jack-William Barotta}
\affiliation{Brown University, Center for Fluid Mechanics and School of Engineering, 184 Hope
St., Providence, Rhode Island 02912, USA
}%
\author{Stuart J. Thomson}%
\affiliation{Brown University, Center for Fluid Mechanics and School of Engineering, 184 Hope
St., Providence, Rhode Island 02912, USA
}%
\affiliation{University of Bristol, Department of Engineering Mathematics, Ada Lovelace Building, University Walk, Bristol, BS8 1TW, UK
}%

\author{Luke F.L. Alventosa}
\affiliation{Brown University, Center for Fluid Mechanics and School of Engineering, 184 Hope
St., Providence, Rhode Island 02912, USA
}%

\author{Maya Lewis}
\affiliation{Brown University, Center for Fluid Mechanics and School of Engineering, 184 Hope
St., Providence, Rhode Island 02912, USA
}%

\author{Daniel M. Harris}
\affiliation{Brown University, Center for Fluid Mechanics and School of Engineering, 184 Hope
St., Providence, Rhode Island 02912, USA
}%
 \email{daniel\_harris3@brown.edu}

\date{\today}

\begin{abstract}
\textbf{Abstract}
\\
When a solid body floats at the interface of a vibrating liquid bath, the relative motion between the object and interface generates outwardly propagating surface waves. It has recently been demonstrated that millimetric objects with fore-aft mass asymmetry generate an associated asymmetric wavefield and consequently self-propel in unidirectional motion. Harnessing this wave-powered mechanism of propulsion, we here demonstrate that chiral objects placed on a vibrating fluid interface are set into steady, yet reversible, rotation, with the angular speed and direction of rotation controlled by the interplay between object geometry and driving parameters.  Scaling laws and a simplified model of the wavefield reveal the underlying physical mechanism of rotation, while collapsing experimental measurements of the angular velocity across parameters. \edit{Leveraging the control over the chiral object's direction of rotation}, we \edit{then demonstrate that a} floating body with an asymmetric mass distribution and chirality can be remotely steered along two-dimensional trajectories via modulation of the driving frequency alone. This accessible and tunable macroscopic system serves as a \textcolor{black}{potential} platform for \textcolor{black}{future explorations of} chiral active and driven matter, and demonstrates a mechanism by which wave-mediated \textcolor{black}{fluid} forces can be manipulated for directed propulsion.
\end{abstract}

\maketitle

\section*{Introduction}
Synthetic \textcolor{black}{self-propelled or active particles} have gained prominence in addressing \textcolor{black}{outstanding} questions in non-equilibrium physics \textcolor{black}{and fluid mechanics} as they offer a relatively higher degree of tunability and parametric control compared to their animate counterparts \cite{palacci2015artificial,bechinger2016active,moran2017phoretic,klotsa2019above,michelin2023self}. \textcolor{black}{While some particles rely on internal} self-actuation, propulsion of otherwise passive particles \edit{can also be realized} by combining external forcing of the environment with symmetry-breaking at the single particle level. For example, the geometry of chiral particles---objects that cannot be superimposed on mirror images of themselves---results in both translational and rotational particle dynamics in myriad systems \cite{liebchen2022chiral} when coupled to a similarly diverse array of driving mechanisms \cite{tsai2005chiral,di2010bacterial,sokolov2010swimming, di2012hydrodynamic,maggi2015micromotors,sabrina2018shape,shields2018supercolloidal,workamp2018symmetry,sabrina2018shape,francois2018rectification,takahashi2021horizontal}. However, in many cases relevant to complex fluids, while symmetry-breaking and activity are minimal ingredients to induce directed motion, pathways to achieving tunable particle dynamics are often obscured owing to an incomplete description of the hydrodynamics underpinning self-propulsion \cite{brooks2019shape}. The ability to design \textcolor{black}{self-propelled} particles informed by an intimate knowledge of the attendant fluid mechanics could lead to advanced control strategies for particle transport and self-organization across scales \cite{aubret2018targeted} and complex navigation of steerable surface-dwelling robots \cite{hu2010water,yuan2012bio,rhee2022surferbot}.

\edit{The vibrating fluid interface is} an emerging \edit{macroscopic} experimental platform \textcolor{black}{for studying self-propulsion and collective dynamics} of particles and \edit{liquid} droplets. While sharing nearly identical hardware with its vibrated granular counterpart and enjoying the same degree of parametric control \cite{tsai2005chiral,narayan2007long,kudrolli2008swarming,scholz2016ratcheting,bar2020self, lowen2020inertial}, the vibrating fluid interface operates on a physically distinct set of principles. 
Specifically, dry granular particles self-propel and interact only through local steric interactions with a solid substrate and neighboring particles, while the underlying physics of propulsion in vibrated granular systems is challenging to model and accurately predict. This contrasts to the vibrating fluid system where \edit{self-propulsion is primarily driven by fluctuations of the fluid interface} and wave-mediated interactions render inter-particle coupling inherently non-local, \edit{phenomena} exemplified by self-propelled liquid droplets \cite{couder2005walking,pucci2015faraday,bush2020hydrodynamic,thomson2020collective,saenz2021emergent} and floating asymmetric solid objects \cite{francois2018rectification, ho2021capillary}. 
Solid particles have two distinct advantages over droplet systems: emergent dynamics may be interrogated over significantly broader parameter regimes and there is \textcolor{black}{significantly more} flexibility in the design space of the particles. The inherent tunability of the vibrating fluid interface platform can therefore be systematically leveraged to control self-propelled vibrated particles, while accompanying models based on hydrodynamics can inform and validate \textcolor{black}{general} design principles.

We herein demonstrate experimentally and theoretically that a chiral particle, henceforth referred to as a spinner, pinned at the interface of a vertically vibrating liquid bath may be set into steady, bidirectional rotational motion thanks to outwardly propagating surface waves generated by the spinner as it oscillates on the fluid surface. We reveal that the direction of induced rotation is not solely prescribed by the spinner's chiral geometry, but from a subtle interplay with the wavelength of the spinner's self-generated wavefield. In particular, we demonstrate that switchable rotation of the spinner is controlled by changing either the size of the spinner or the external driving frequency of the liquid bath, both of which lead to a change in the geometry of the spinner's self-generated wavefield. Informed by the \textcolor{black}{relevant} aspects of the fluid system, scaling arguments rationalize the experimentally measured angular velocities for various sizes of spinners and driving parameters, while a simple theoretical model of the stresses that accrue through the spinner's self-generated surface waves illuminates the hydrodynamic origins of the reversibility of the spinner rotation. With the insight gained from \textcolor{black}{our} understanding of the chiral spinner, we design a chiral active particle with both mass- and geometric-asymmetry that can be remotely steered along the fluid interface via modulation of the external driving frequency alone. We propose that the relative characteristic length scale of the particle size to the background driving field may serve as a control mechanism for achieving bespoke particle trajectories in \textcolor{black}{more general contexts}.

\section*{Results}

\subsection*{Steady Rotation of Chiral Spinners}
\begin{figure*}[htb!]
    \centering
    \includegraphics[width = 18cm ]{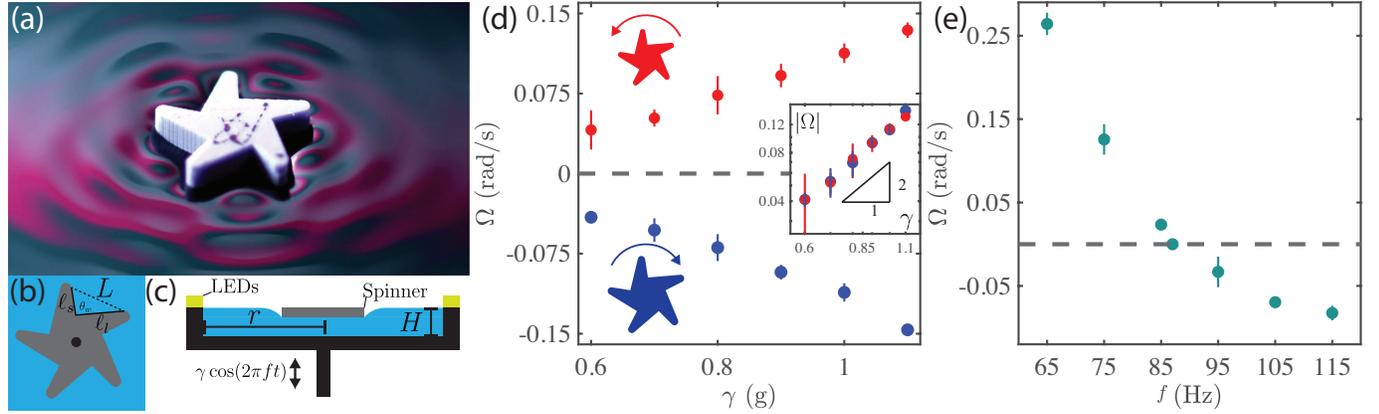}
    \caption{Illustration of the experimental setup and angular velocity data. (a) Oblique perspective of a chiral spinner on the surface of a vibrating fluid bath. The self-generated capillary wavefield of the spinner is visualized by the distortions of a reflected color pattern \cite{harris2017visualization}. 
    (b) A schematic of the spinner geometry: each wedge consists of a long arm of length $\ell_l$ and short arm of length $\ell_s$, separated by an angle $\theta_w = 94.4\pi/180$. All spinners are geometrically similar and are characterized by the opening length $L$. (c) A schematic of the experimental set-up: a chiral spinner rests on the surface of an illuminated circular liquid bath of height $H=5.7$ mm and radius $r=50$ mm. The bath is oscillated vertically at a frequency $f$ with maximum acceleration amplitude $\gamma$. (d) Angular velocity of a large ($L=6.3$ mm, blue) spinner and small ($L=4.6$ mm, red) spinner as function of $\gamma$ at fixed frequency $f=95$ Hz for $\gamma/g$ = (0.6, 0.7, \ldots, 1.1). Inset: a square dependence of the angular speed on the driving acceleration for both sizes of spinner is consistent with the physics of wave radiation stress. (e) For a fixed value of $\gamma = 1$ g, the angular velocity of a spinner ($L = 5.5$ mm) decreases monotonically as function of $f$ from $f = 65$ Hz to the flipping frequency $f^* \approx 87$ Hz. Thereafter, the angular velocity switches sign. The error bars in (d) and (e) represent one standard deviation in the angular velocity over at least four trials, \textcolor{black}{and in some cases are smaller than the marker size itself.}}
    \label{fig:fig1}
\end{figure*}

To systematically investigate the fluid mechanics regulating the rotation of the spinners, we designed hydrophobic, geometrically similar five-pointed chiral stars whose center of mass coincides with their geometric center (Fig.\ \ref{fig:fig1}(a) and (b)). A complete description of the spinner fabrication process can be found in the Methods section. \edit{As depicted in Fig.\ \ref{fig:fig1}(b)}, chirality is induced by five wedges separating the points of the star, parameterized by a wedge opening angle of $\theta_w = 94.4\pi/180$ and short- and long-arm lengths of $\ell_s$ and $\ell_l$, respectively, with $\ell_l/\ell_s=1.7$. Geometric similarity thus allows the spinners to be characterized by a single length scale, chosen here to be the wedge opening length $L$. The hydrophobic chiral stars are then placed on the surface of a liquid bath consisting of a water-glycerol mixture, which in turn is mounted on an in-house electromagnetic shaker \cite{harris2015generating} (Fig.\ \ref{fig:fig1}(c)) vibrating vertically with acceleration $\Gamma(t) = \gamma\cos2 \pi f t$, where $\gamma$ is the maximum vertical acceleration, $f$ is the vibration frequency, and $t$ is time.  All experiments reported herein were conducted below the Faraday threshold of the fluid, and hence the free surface remains flat in the absence of the spinners.

A spinner residing on the vibrating fluid surface quickly established an extended capillary wavefield and reached a steady rotation rate with angular velocity $\Omega$.  All spinners used in the experiments had the same chirality, with positive and negative $\Omega$ corresponding to counterclockwise (CCW) and clockwise (CW) rotation, respectively. Focusing in the first instance on two spinners of different sizes ($L$ = 4.6 mm and 6.3 mm) and fixing the frequency $f = 95$ Hz, we found that in both cases $\Omega$ varies monotonically with $\gamma$,  \textcolor{black}{with a best-fit power law yielding $\Omega \sim \gamma^{1.99 \pm 0.06}$ and $\Omega \sim \gamma^{2.20 \pm 0.10}$ for $L = 4.6$ mm and $L = 6.3$ mm, respectively} (Fig.\ \ref{fig:fig1}(d)). \textcolor{black}{Due to experimental limitations the best-fit power-law scalings for $\Omega$ are determined in a regime where the variation in $\gamma$ is less than an order of magnitude. Nevertheless, the suggested} square dependence of the angular velocity on the driving \textcolor{black}{amplitude} at fixed frequency is consistent with the physics of wave radiation stress \cite{longuet1964radiation, ho2021capillary}. While the two sizes of spinners considered here possess the same chirality and differ only by the length scale $L$, the smaller spinner was observed to rotate in a CCW direction, the larger spinner in a CW direction. This phenomenon may be further interrogated by considering the influence of varying driving frequency on the spinner rotation. 

\begin{figure*}[htb!]
    \centering
    \includegraphics[width = 18cm]{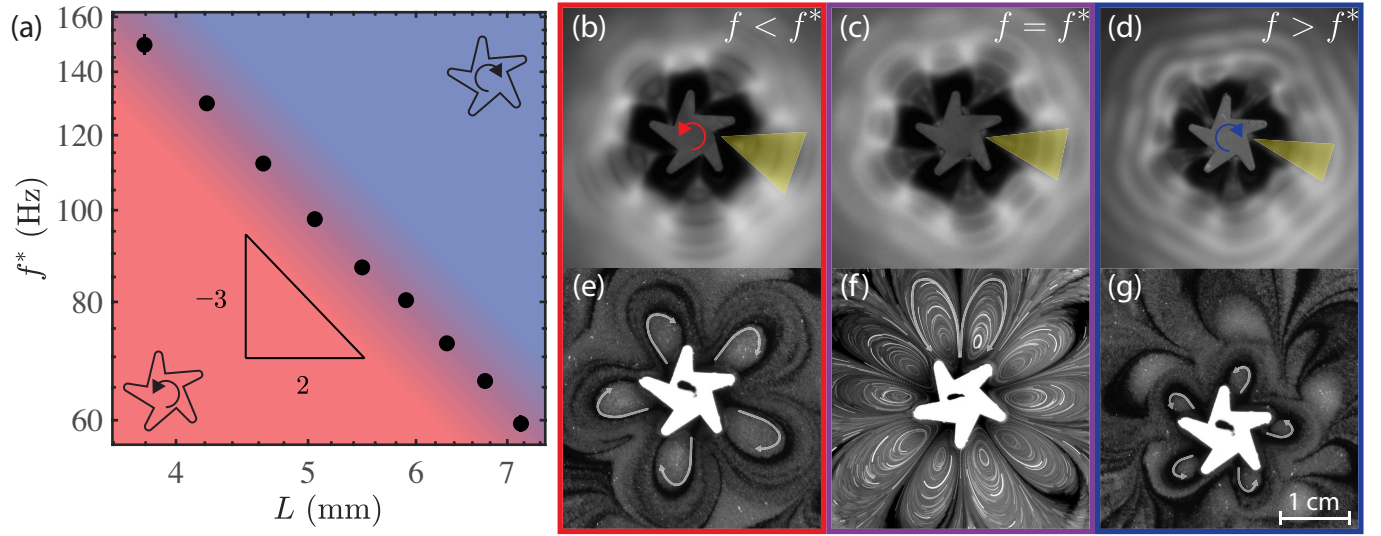}
    \caption{Bidirectional rotation and the flipping frequency.  (a) Plot of the flipping frequency $f^*$ at which $\Omega = 0$ for nine different spinner sizes in the range $L=3.8-7.2$ mm. For $f<f^*$ the spinner rotates CCW (red region) and CW for for $f>f^*$ (blue region). (b-d) We use a shadowgraph imaging technique to visualize the wave field generated by the oscillating spinner and in (e-g) distributing tracer particles at the interface reveals qualitatively similar characteristic streaming-flow patterns for a typical spinner of size $L = 6.3$ mm across the range of frequencies considered in experiments. In (b--d), constructive interference of waves propagating from each arm of the spinner gives rise to rays (indicated by yellow triangles) centered at the vertex of each of the five wedges. As the frequency increases, the bands of constructive interference narrow and the flow structure changes from having (e) a single bound vortex attached to the long arm of the spinner when $f < f^*$ to (f) two closed vortices in each wedge when $f = f^*$ and then finally to (g) a single bound vortex attached to the short arm of the spinner for $f > f^*$. \textcolor{black}{See Supplementary Videos 1--4 for real-time movies of the bidirectional rotation.} The error bars in (a) represent one standard deviation in the flipping frequency over six trials, \textcolor{black}{and in some cases are smaller than the marker size itself.}}
    \label{fig:fig2}
\end{figure*}

\subsection*{Bidirectional Rotation and the Flipping Frequency}

Fixing both the spinner size ($L=5.5$ mm) and driving acceleration $\gamma=1$g, a frequency sweep was performed, increasing $f$ incrementally from 65 to 115 Hz. In this case $\Omega$ decreases monotonically to zero, eventually changing sign at the flipping frequency $f = f^{*}$, whereafter the spinner rotates in the opposite direction (Fig.\ \ref{fig:fig1}(e)). Further, we found that $f^{*}$ is independent of $\gamma$ for a given spinner and hence the strength of the vibrational forcing does not influence the direction of rotation, only the angular velocity. \textcolor{black}{The value of $f^*$ was measured for nine different spinner sizes of varying $L$, resulting in an inverse relationship between the two parameters} (Fig. \ref{fig:fig2}(a)). In each case, $f^{*}$ was found to be independent of $\gamma$. 
The observed dependence of $f^{*}$ on $L$ may be rationalized by considering the relevant physical lengthscales present. The propagating waves excited by the spinner as it oscillates on the fluid surface are synchronous with the forcing frequency and are governed by the dispersion relation for deep-fluid capillary waves \cite{de2004capillarity,lamb1924hydrodynamics}, namely $\omega^2 = \sigma k^3/\rho$, where $\omega = 2\pi f$ is the angular frequency, $k = 2\pi/\lambda$ is the wavenumber, $\sigma$ is the surface tension of the fluid and $\rho$ is fluid density. (The gravitational component of the dispersion is negligible due to the relatively large frequencies considered here, namely $f \gtrsim 30$ Hz.) Hence, the capillary wavelength $\lambda \sim f^{-2/3}$.  Since the spinners are characterized by a single length scale $L$, we anticipate the flipping frequency to occur when $\lambda$ and $L$ become comparable, leading to $f^{*}\sim L^{-3/2}$. A best-fit power law to the data in Fig.\ \ref{fig:fig2}(a) \textcolor{black}{is consistent with} this prediction yielding $f^* \sim L^{-1.44\pm 0.01}$. \textcolor{black}{(Similar to the scaling relationship between $\Omega$ and $\gamma$, due to experimental limitations the best-fit power-law scaling for $f^*$ is derived from data that spans less than an order of magnitude in $L$.)} This result suggests that the rotation direction is defined by the geometry of the wavefield, which is set by the shape and size of the wave source (spinner) and the frequency-dependent wavelength of the spinner's self-generated capillary wavefield. In summary, the direction of spinner rotation can be controlled by either the spinner size or the driving frequency, $f$.

\subsection*{Parametric Dependence of Rotation Speed}
Our experiments demonstrate that the amplitude (set by $\gamma$) and wavelength (set by $f$) of the spinner's self-generated wavefield (visualized in Fig.\ \ref{fig:fig2}(b-d)) are responsible for the speed and direction of spinner rotation, respectively. We now synthesize these observations to rationalize the rotation speed of the spinners in terms of the physical parameters of the fluid system. Specifically, we consider a dominant balance of the driving torque arising from wave radiation stress, which we denote $\tau_{\lambda}$, with the resisting torque $\tau_{\mu}$ due to viscous friction of a locally fully developed shear flow beneath the spinner as it rotates on the fluid surface.

\begin{figure}[htb!]
    \centering
    \includegraphics[width = 9cm]{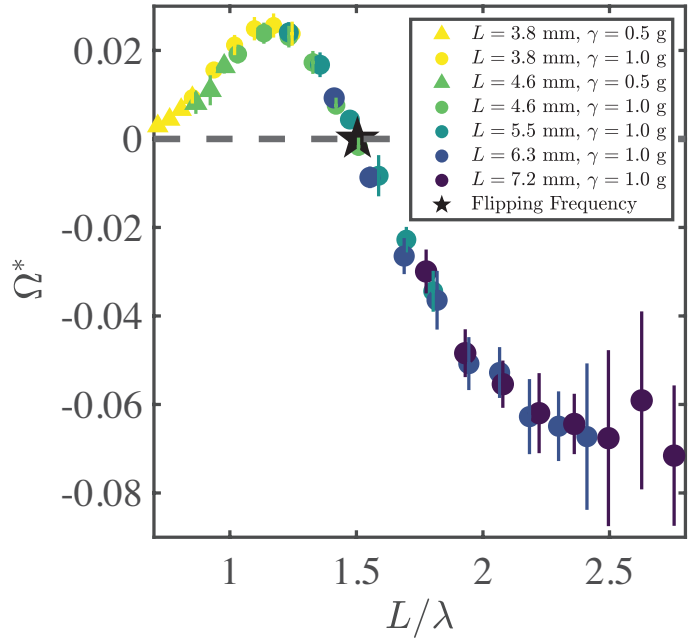}
    \caption{The dimensionless angular velocity as a function of the dimensionless spinner size. The data from spinners for different combinations of frequencies, forcing \textcolor{black}{amplitudes}, and sizes all collapse along a non-monotonic curve $F(L/\lambda)$. The flipping frequency (star) occurs at $L/\lambda \approx 1.5$. The error bars represent one standard deviation in $\Omega^*$ which result from standard multivariate error propagation, using uncertainties in the angular velocity and fluid parameters described in Methods.}
    \label{fig:fig3}
\end{figure}

In the capillary regime, the force per unit length from wave radiation stress \cite{longuet1953mass, longuet1964radiation} scales with $\sigma(kA)^2$ and hence $\tau_{\lambda}\sim\sigma(kA)^2L^2F(L/\lambda)$, where the unknown function $F(L/\lambda)$ embeds the details of the wavefield geometry and depends only on the ratio $L/\lambda$ since the spinner shape is fixed. For a fluid of depth $H$ and dynamic viscosity $\mu$, we expect viscous stress along the base of the spinner to scale with \textcolor{black}{$\mu\Omega L/H$} and thus \textcolor{black}{$\tau_{\mu}\sim\mu \Omega L^4/ H$.}  Balancing $\tau_{\lambda}\sim \tau_{\mu}$ and assuming $A\sim\gamma / \omega^2$, we find a dimensionless angular velocity
\begin{equation}
\label{eqn:omegaStar}
    \Omega^* = \frac{ \Omega}{\left[ \frac{H\rho^2 \gamma^2}{\mu \sigma} \frac{ 1}{k^4L^2} \right]} \sim F(L / \lambda).
\end{equation}
In Fig.\ \ref{fig:fig3} we plot experimental data for $\Omega^{*}$ obtained for a family of different values of $L$, $\lambda$ \textcolor{black}{(by changing $f$)}, and $\gamma$, with the data collapsing along a non-monotonic curve $F(L/ \lambda)$ that passes through zero at approximately $L/\lambda\approx1.5$. 
Although beyond the scope of the present work, we note that the predicted dependencies of the rotation on the fluid parameters ($\mu$, $\sigma$, $\rho$, and $H$) in the scaling law equation \eqref{eqn:omegaStar} remain to be verified. \textcolor{black}{However, the successful collapse does confirm the proposed scaling on the geometric and driving parameters $L$, $f$, and $\gamma$.}

It is well-known that wave fluctuations give rise to quasi-two-dimensional surface streaming flows \cite{longuet1953mass, filatov2016nonlinear, francois2013inverse, parfenyev2020large}, suggesting a second candidate driving mechanism alongside wave-radiation stress (Fig.\ \ref{fig:fig2}(e-g)). However, while the surface streaming flows are a clear visual marker of the wavefield and direction of rotation, they represent only a higher-order contribution to the momentum balance \textcolor{black}{as the waves are of relatively small amplitude in the present work, specifically $A/\lambda \lesssim 0.1$}. Surface flows generated from small-amplitude surface waves generally have a characteristic flow velocity $u \sim (kA)^2$ associated with Stokes' drift \cite{francois2017wave} and Eulerian flow contributions \cite{filatov2016nonlinear}. Thus the momentum flux associated with the flow field is expected to follow the scaling $\rho u^2 \sim \rho(kA)^4$, at higher order in wave amplitude than wave radiation stresses. We therefore conclude that wave-radiation stress, and not streaming-flows, are the leading-order cause of spinner rotation.

\subsection*{Wave Propulsion Allows for Bidirectional Spinning}

While equation \eqref{eqn:omegaStar} for $\Omega^*$ defines the dependence of the rotation rate on a single function and highlights wave-radiation stress as the primary driver of rotation, it does not yield insight into the non-monotonicity of $F(L / \lambda)$, nor the mechanism that enables the spinners to change their direction of rotation. We are thus prompted to consider a simple model of the spinner's self-generated capillary wavefield to analyze the role of the wavefield geometry on the direction of rotation. Modeling each arm of the spinner as a wavemaker emitting small-amplitude, unidirectional, planar capillary-waves with amplitude $A$, the wave profile within a wedge of the spinner is the superposition of two crossed waves \cite{parfenyev2020large,bliokh2022field, poplevin2021formation}, namely
\begin{equation}
    h(x,y,t) = A\cos(ky\cos\alpha+ kx\sin\alpha - \omega t) + A\cos(ky\cos\alpha - kx\sin\alpha - \omega t),
        \label{eqn:eqncross}
\end{equation}
where $\alpha = (\pi - \theta_w)/2$. 

For unidirectional capillary waves propagating in the $y$-direction, the wave radiation stress is given by $S_{yy} = 3\sigma k^2\langle h^2 \rangle/4$, where $\langle h^2 \rangle = 2A^2 \cos^2\left(kx\sin\alpha\right)$ is the time-averaged \textcolor{black}{squared} amplitude \cite{longuet1964radiation}. (We note the term radiation stress is used to remain consistent with the extant literature, even though $S_{yy}$ has units force-per-unit-length.) Placing the vertex of the wedge at the origin $(x,y) = (0,0)$, the torque produced by the wave radiation stresses, $\tau_{\lambda}$, from the five wedges can be represented by the line integral over the short and long arms, namely
\begin{equation}
\label{eqn:theoryTorque}
    \tau_{\lambda} = -5\int_{\ell_l + \ell_s} \mathbf{r} \times \left( S_{yy} \mathbf{n}\right)\ \text{d}s,
\end{equation}
where $\mathbf{n}$ is the unit outward-facing normal vector along the perimeter of the wedge and $\mathbf{r}$ is the radial coordinate, relative to the geometric center of the spinner.

\begin{figure}[htb!]
    \centering
    \includegraphics[width = 9cm]{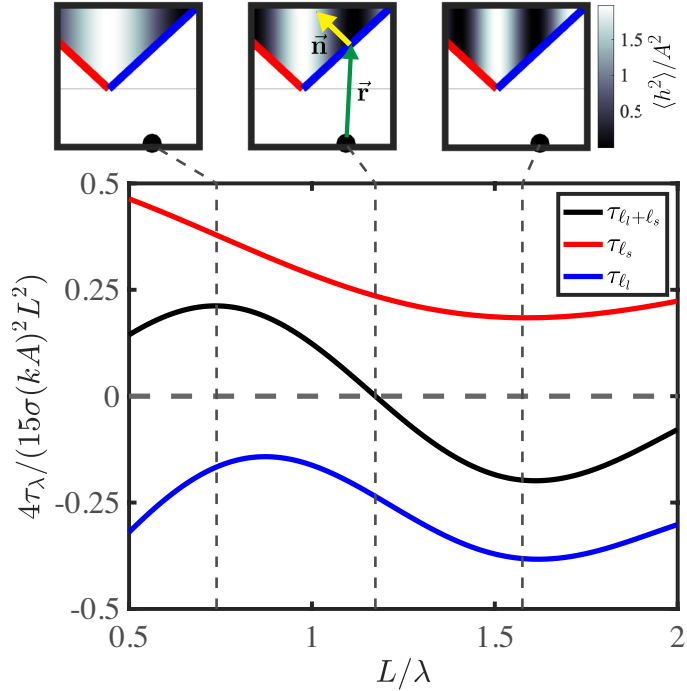}
    \caption{Theoretical dimensionless torque as a function of the dimensionless spinner size ($L/\lambda$). The black curve represents the sum of the contributions from the short arm (red curve) and long arm (blue curve).  For smaller sizes of spinner or low vibrational frequency, the net torque is positive and exhibits a local maximum before eventually flipping sign as the size of the spinner or driving frequency is increased. The time-average square wavefields in the wedge as predicted by the crossed-wave model \eqref{eqn:eqncross} are shown along the top for three representative cases. The black dot in the lower corner of each box denotes the geometric center of the spinner.}
    \label{fig:fig4}
\end{figure}

The dimensionless torque computed using equation \eqref{eqn:theoryTorque} is plotted in Fig.\ \ref{fig:fig4} and captures the salient features described by $F(L/\lambda)$ in Fig.\ \ref{fig:fig3}. Specifically, for smaller values of $L$ or at low frequency (longer wavelengths), the spinners are predicted to rotate in a CCW direction, switching their direction of rotation when $L\sim\lambda$, and rotating in a CW direction thereafter. While of the same order of magnitude, the discrepancy between experiments and our theoretical prediction of the critical value of $L/\lambda$ for which switching occurs likely rests in the simplifying assumptions made in posing the model for the wavefield $h$, which neglects wave dissipation and finite-size effects of the spinner. Furthermore, our simple model predicts that for larger $L/ \lambda$ the spinner switches its direction of rotation direction again, a feature not observed in experiments. At high frequencies ($L/\lambda\gg 1$) finite-size effects and viscous dissipation are more prominent, rendering the model more difficult to justify in this regime.

The crossed-wave profile, $h$, yields a length-scale $\pi /( k\sin\alpha)$ of interference bands of constructive/destructive interference \cite{parfenyev2020large} whose width \textcolor{black}{decreases} as the forcing frequency increases, similar to the experimentally observed bands of constructive interference centered on a ray emanating from the wedge vertex in Figs. \ref{fig:fig2}(b-d). The corresponding \edit{time-averaged square} wavefields at different $L/\lambda$ predicted by the crossed-wave model are presented at the top of Figure \ref{fig:fig4}.  For small $L/\lambda$, only a single ray is predicted, emanating from the center of the wedge.  Thus, the majority of the wave stress is isolated to the region near wedge corner and is nearly symmetric.  However, the net torque contributions are not symmetric as a result of the geometry of the spinner: the small arm has a larger mechanical advantage (longer lever arm) on the overall spinner rotation and hence, consistent with our computation of $\tau_{\lambda}$, we anticipate a net positive wave torque in this regime.  For larger $L/\lambda$, a second ray emerges at the tip of the long arm of the spinner, allowing the long arm to overcome the torque from the small arm and hence the spinner rotates in the opposite direction.  

In spite of its limitations, the simplified model of the \textcolor{black}{spinner’s} self-generated wavefield considered here adequately demonstrates that switchable rotation is achieved thanks to the subtle interplay between different-size effective lever arms of the spinner, chirality, and the spatially structured wave forcing resulting from the wedge-shaped wavemaker.  \textcolor{black}{In preliminary experiments, chiral spinners with other wedge angles, number of arms, and arm shapes also exhibited bidirectional spinning and a non-monotonic frequency response.  Thus the key qualitative behaviors detailed in the present work are not peculiar to the specific geometry considered, however, the quantitative details will inevitably differ.}  

\subsection*{Frequency-steerable Chiral Active Particles}

\begin{figure}[htb!]
    \centering
    \includegraphics[width = 9cm]{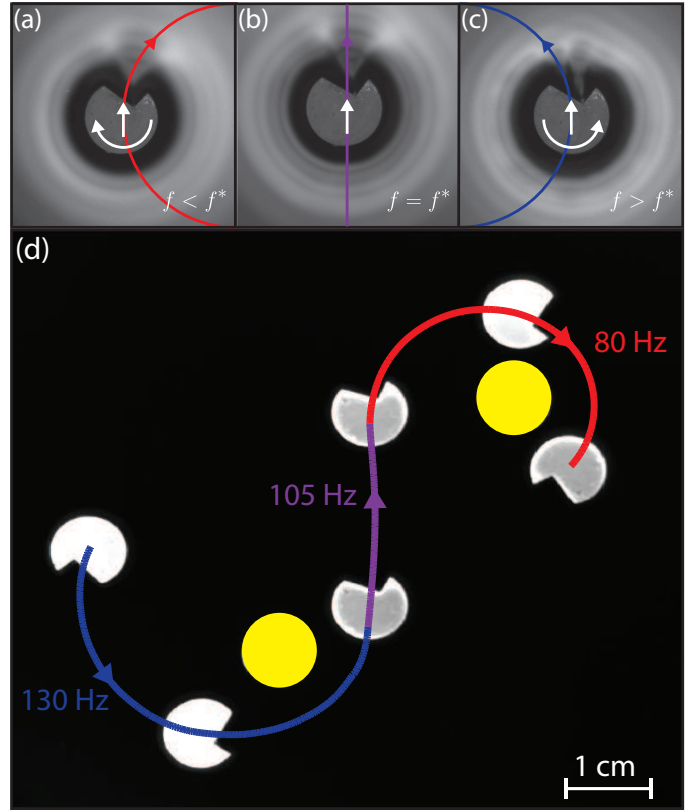}
    \caption{Remote steering of a ``Pac-Man''-shaped chiral active particle (CAP) via modulation of the driving frequency. The CAP translates in the direction of the wedge opening, but simultaneously rotates with a frequency-dependent direction. (a) For $f< f^*$, the CAP tracks a clockwise arc. (b) When $f = f^*$ the CAP follows a straight line. (c) For $f > f^*$, the CAP executes a counter-clockwise arc. (d) The driving frequency can be tuned in real time to force the CAP along a bespoke trajectory. Here three frequencies are used to successfully navigate the CAP between (yellow) circular markers placed on the bottom of the fluid bath. \textcolor{black}{See Supplementary Video 5} for a movie of the navigation.}
    \label{fig:fig5}
\end{figure}

Informed by our understanding of the physics driving the spinner rotation, we extend the functionality of these particles by conceiving of more general geometries to perform specific tasks. We designed a chiral active particle (CAP) by cutting a single wedge with opening length $L = 4.8$ mm from a disc of diameter $8.1$ mm, resulting in a ``Pac-Man'' shape (Fig.\ \ref{fig:fig5}). If the wedge is symmetric ($\ell_s = \ell_l$), mass asymmetry leads to self-propelled, rectilinear motion in the direction of the wedge \cite{ho2021capillary,rhee2022surferbot}. However, an asymmetric wedge induces an additional rotational motion from a net torque, guiding the CAP along smooth arcs whose direction may be prescribed by changing the driving frequency of the bath, $f$ (Fig.\ \ref{fig:fig5}(a)--(c)). For the particular shape used, we found that frequencies near $105$ Hz allowed the CAP to navigate along a straight line indicating that this frequency is analogous to the flipping frequency $f^*$ described for the spinners. When $f < f^* \approx 105$ Hz, a clockwise rotation of the CAP occurs along with the steady translation and the object executes a right turn, while for $f > f^*$ a counter-clockwise rotation occurs allowing the object to make a left turn.

To test the maneuverability of the CAP, we designed an obstacle course defined by two circular markers placed at the bottom of the fluid bath. We were able to successfully navigate the CAP through the course by changing the vibration frequency in real time between 130 Hz (left turn), 105 Hz (straight), and 80 Hz (right turn) (Figure \ref{fig:fig5}(d)). \textcolor{black}{We note that the precise turning radius of the object is a function of the driving frequency as well as particle geometry, likely related to the details of non-monotonic curve $F(L/\lambda)$ and the balance between rotational to translational propulsion generated by the CAP.}
\section*{Discussion}

Recently, chiral particle designs have been leveraged to convert a wide variety of background energy sources into directed rotation of an active particle.  Of these systems, very few have \textcolor{black}{exhibited}  the possibility of bidirectional rotation.  Recent examples of microscopic spinner systems exhibiting bidirectional rotation include the use of ultrasound \cite{sabrina2018shape} or electrohydrodynamic flows (EHD) \cite{shields2018supercolloidal}, both of which demonstrated rotation direction changes via modulation of the driving frequency.  Despite these recent advances, our system is the first to clearly demonstrate that the relative characteristic length scale of the particle size to the background driving field is the primary ingredient enabling bidirectional rotation in a periodically driven system.  Our findings \textcolor{black}{may potentially} inform future investigations at the microscale, including the possibility of frequency-steerable colloidal particles.

More generally, many active and non-equilibrium systems exhibit an intrinsic length scale selection \cite{sokolov2012physical, lee2017fluctuation, heinonen2022emergent}.  Thus, for embedded passive objects or confinement at comparable length scales, different qualitative behaviors can be anticipated depending on the ratio of the lengths, as in the present system.  For instance, when active bacterial suspensions are confined at length scales comparable to the typical bacterial vortex size, they exhibit highly geometry-dependent rheological behaviors \cite{slomka2017geometry,liu2019rheology} and rectification of otherwise irregular flow patterns \cite{wioland2013confinement,wioland2016directed}.  Our study may thus contribute to the growing understanding of controlling forces in non-equilibrium systems \cite{lee2017fluctuation}, an essential step towards advancing and unlocking applications.

\subsection*{Applications to Synchronization and Collective Behaviors}

\textcolor{black}{While the present work focuses on understanding the self-propulsion of a single particle}, multiple self-propelled objects can interact dynamically at long range through their extended wave and flow fields \cite{ho2021capillary}.  In general, recent work has just begun to reveal collective behaviors of self-propelled particles that interact through deformable substrates, both on elastic \cite{li2022field} and fluid interfaces \cite{thomson2020collective,ho2021capillary,saenz2021emergent}.  Furthermore, \textcolor{black}{the capillary spinner} system represents a highly controllable ``wet'' active system with non-negligible inertial effects in both the particle and fluid dynamics. For example, the Reynolds number associated with the spinner rotation on the surface of the fluid is $\text{Re} = \rho \Omega L^2 / \mu \sim \mathcal{O}(10)$. This intermediate Reynolds number regime is largely unexplored from both non-equilibrium physics and hydrodynamics perspectives \cite{bechinger2016active,klotsa2019above}. \textcolor{black}{Ongoing work is exploring} pairs \textcolor{black}{and} collections of spinners that exhibit rich collective interactions such as synchronization \cite{strogatz2000kuramoto} due to their long-range hydrodynamic interactions. Since each spinner has a well defined phase and the freedom to move laterally along the fluid interface, our macroscopic system embodies the key ingredients of so-called ``swarmalators'': mobile oscillators that can both synchronize and swarm \cite{o2017oscillators}.
 
\subsection*{Applications to Robotics}

Over the past decade or so, considerable engineering effort has been devoted to the development of ``microrobots'' (typically at millimeter or centimeter scales) that self-propel along the air-water interface, motivated by applications such as environmental monitoring \cite{hu2010water,yuan2012bio}. A vast range of propulsion strategies for such devices have been shown to be possible, many of which are inspired by biological counterparts found in nature.  Of most relevance to the present study, self-generated propagating capillary waves can be used to drive propulsion of an untethered vibrating robotic device (the ``SurferBot'') \cite{rhee2022surferbot}, inspired by a recently documented survival strategy of a water-bound honeybee \cite{roh2019honeybees}.  \textcolor{black}{Despite sharing the same physical wave-propulsion mechanism as the capillary spinners presented herein, the robotic device does not rely on an external vibration source (e.g. vibrating bath) but rather is the source of its own vibrations via} simple and inexpensive onboard electronics (vibration and power source). \textcolor{black}{Other recent works on a similar robotic device have demonstrated that steering can be enabled by including a second independent onboard vibration source \cite{wang2022miniature}.  As an alternative, the present findings suggest that} the incorporation of a chiral \textcolor{black}{base shape} and \textcolor{black}{adjustable} driving frequency \textcolor{black}{might} represent \textcolor{black}{another viable mechanism to realize} real-time steering of such a device \textcolor{black}{with fewer components}. \textcolor{black}{It has been suggested that control and scale-down of active particles or robots may be advanced by increased exploitation of their physical interactions rather than solely relying on advancing the onboard hardware or computational capabilities \cite{li2021programming}.}  The design of other interfacial microbots that rely on on-board vibration or oscillating contact lines may similarly be informed by the present study \cite{lee2019milli,jiang2020triboelectric}.

\subsection*{Applications to Surface Particle Manipulation}

While the objects and flow structures presented here exist on millimeter to centimeter scales, considerable recent work has demonstrated that macroscale static or quasi-static interfacial curvature can be used as an effective ``field'' to drive assembly \cite{liu2018capillary} and manipulate \cite{zeng20223d} microscopic particles.  In addition, capillary waves have been shown to \textcolor{black}{organize and cluster} particle distributions at an interface \cite{falkovich2005floater} and stabilize capillary attraction in certain regimes \cite{de2018capillary,ho2021capillary}.  Furthermore, secondary velocity fields arising from oscillatory surface flows have been applied to manipulate particles at a distance using only fluid flow \cite{punzmann2014generation,francois2017wave,chavarria2018geometrical,abella2020spatio}, a mechanism that is even potentially exploited by some organisms for feeding \cite{joo2020freshwater}.  With both the spinner and the ``Pac-Man'' geometries explored here, extended wavefields and vortical flows are observed at the interface that emanate from the wedges as seen in Fig.\ \ref{fig:fig2}(e-g), with both the vortex size and topology changing with the driving frequency.  Micron-scale tracer particles are observed to become trapped in closed streamlines that move with the object and can be released via changes in the remote driving parameters.  Ongoing work is focused on exploring the extent to which microscopic surface particles can be transported, assembled, sorted, or otherwise manipulated using these \textcolor{black}{tunable} wave and flow sources.

 \section*{Methods} 
 \subsection*{Fabrication of the Chiral Spinners}
 The spinners were fabricated using OOMOO 30, a commercially available two-part (Part A and B) silicone rubber compound. Working quickly due to the 20 to 30 minute pot-life of OOMOO 30, both Part A and Part B were stirred vigorously for 10 seconds with clean wooden sticks. 60 milliliters of Part A and 60 milliliters (mL) of Part B were then poured into a clean disposable plastic cup and mixed for 20 seconds to create a mixture of 1:1 by volume. 

 Once well-mixed, the mixture was poured into 3D-printed (Elegoo Mars 2 Pro) molds designed in Fusion 360. Rounded corners at the five points of the star-shaped spinner were added to help maintain the pinned contact line throughout the course of the experiments.  (The CAD files for the spinner design are included as part of the Supplementary Material.) To pop larger bubbles that formed during the mixing process and to allow the mixture to make its way into some of the smaller crevices in the molds (the vertices of the star, for example), the mixture was poured from a height of around 0.5 meters. Excess compound was then scraped off the top of the mold using a straight-edge razor blade. 

 Finally, the molds containing the mixture were placed in a vacuum chamber for a total of 60 seconds in four 15 second intervals to remove any smaller air bubbles present in the mixture. The surface of the molds were then scraped firmly one final time leaving little-to-no excess compound. The liquid mixture was then cured at room temperature for at least 6 hours, typically overnight. Once cured, the 2mm-thick spinners are readily removed from the mold. To ensure repeatability across batches of spinners made, the flipping frequency, $f^*$, was tested for each independent preparation and no systematic differences were observed. Any spinners that had visible bubbles were discarded and not used for data acquisition. An identical fabrication process was performed to manufacture the ``Pac-Man''-shaped chiral active particles (CAPs). The cured silicone-rubber compound is naturally hydrophobic. Thus, surface tension and symmetric mass distribution allow the spinners to rest without tilt on the surface of a liquid bath consisting of a water-glycerol mixture, with a pinned contact line along the lower perimeter of the spinner or CAP. 

 \subsection*{Experimental Details}
 The fluid used in all experiments was a water-glycerol mixture, with a density of $\rho = 1138.9 \pm 0.3$ kgm$^{-3}$ that was measured daily with a density meter (Anton Paar DMA 35A).  The surface tension of the water-glycerol mixture \cite{glycerine1963physical} was $\sigma = 0.068 \pm 0.001$ Nm$^{-1}$, while the dynamic viscosity\cite{cheng2008formula} was $\mu = 0.00799 \pm 0.00040$ kgm$^{-1}$s$^{-1}$. $\rho$, $\sigma$ and $\mu$ were consistent for all experiments. For each experiment, $45 \pm 1$ mL of the water-glycerol solution was measured using a graduated cylinder and poured into a circular bath (depth $H=5.7$ mm and diameter 100 mm) constructed from laser-cut sheets of acrylic. The fluid depth coincides with the thickness of the acrylic sheet to avoid the formation of a meniscus that would otherwise generate small capillary waves emanating from the boundaries of the bath. 

 Prior to deposition on the fluid surface, each spinner was rinsed with deionized water and dried with a dry cleaning wipe (Kimtech Science). A clean pair of tweezers was used to place the spinner on the fluid surface, with care being taken to ensure the contact line was pinned along the bottom periphery of the spinner. In the absence of vertical vibration of the bath, the spinner and fluid remained stationary.

 The bath is mounted atop a rigid aluminum mounting platform and driven vertically using an electrodynamic modal shaker (Modal Shop, Model 2025E).  The mounting platform is connected to the shaft of a linear air bearing that interfaces with the shaker via a long flexible stinger, ensuring uniaxial motion at the platform \cite{harris2015generating}.  For all frequencies and amplitudes explored in the present work, the vertical vibration amplitude was uniform to within 1\%. The acceleration amplitude was monitored using two accelerometers (PCB Piezotronics, Model 352C65) mounted diametrically opposed on the driving platform, and a control loop maintained a constant driving amplitude.  The entire vibration assembly is mounted on a vibration isolation table (ThorLabs ScienceDesk SDA75120) to minimize the influence of ambient vibrations.

 \subsection*{Tracking Rotational Motion of the Spinners}
 The spinners were filmed at 20 frames-per-seconds (fps) using an overheard camera (Allied Vision, Mako U-130B) with a macro lens placed 1 meter above and normal to the bath surface. An LED-light array surrounding the bath provides a sharp contrast between the spinner (white) and the bottom of the fluid bath (black), allowing for accurate tracking of the spinner. Using the Image Processing Toolbox in MATLAB, the video frames are then processed with $\texttt{imregtform}$ and \texttt{tformType} set to \texttt{rigid}, thus accounting for both translation and rotation of the spinner. The routine operates between consecutive frames to obtain both the translation of the spinner and the angle of rotation $\Delta \theta$. The value of $\Delta \theta$ is calculated by taking the inverse tangent between the components the rotation matrix block of the outputted affine matrix. To avoid the tracking algorithm jumping to different arms during post-processing, we take $\text{mod}(\Delta \theta, 2\pi/5)$. The angular velocity of the spinner is then $\Omega = \Delta \theta\times 20\ \text{fps}$. A random dot pattern was added using India Black ink to enable absolute angular position tracking despite the shape's 5-fold symmetry.

\section*{Data Availability}
The authors declare that all data supporting the findings of this study are
available within the paper and its supplementary information files.

\section*{Acknowledgements} 
We thank Robert Hunt for many useful discussions as well as assistance with setting up the initial iterations of the spinner tracking code used for processing videos, Giuseppe Pucci for assistance in the setup for the shadowgraph visualizations, and Nilgun Sungar for useful discussions. J.-W.B is supported by the Department of Defense through the National Defense Science \& Engineering Graduate (NDSEG) Fellowship Program.

\section*{Author Contributions}
J.-W.B., S.J.T., and D.M.H. designed research and wrote the paper; J.-W.B. and D.M.H. analyzed data and developed models; all authors performed research, discussed the results, commented on the manuscript and gave final approval for publication, agreeing to each be held accountable for the work performed therein; D.M.H. supervised the project and secured funding.


\bibliography{spinners_refs}
\end{document}